\documentclass[aps,twocolumn,showpacs,preprintnumbers,amsmath,amssymb]{revtex4}
\usepackage{graphicx}% Include figure files
\usepackage{bm}% bold math

\begin{document}

%\draft
\title{Possible symmetries of the superconducting order parameter in
a hexagonal ferromagnet}
\author{K. V. Samokhin}
\affiliation{Department of Physics, Brock University,
St.Catharines, Ontario, Canada L2S 3A1} \date{\today}

\begin{abstract}
We study the order parameter symmetry in a hexagonal crystal with
co-existing superconductivity and ferromagnetism. An experimental
example is provided by carbon-based materials, such as
graphite-sulfur composites, in which an evidence of such
co-existence has been recently discovered. The presence of a
non-zero magnetization in the normal phase brings about
considerable changes in the symmetry classification of
superconducting states, compared to the non-magnetic case.
\end{abstract}


\maketitle

The recent discoveries of superconductivity co-existing with
ferromagnetism in such transition element compounds as ZrZn$_2$
\cite{zrzn2-exp} and UGe$_2$ \cite{uge2-exp} have called for a
revision of our views on the interplay of the two phenomena.
Symmetry considerations can help us to identify the possible order
parameters, even if the microscopic mechanism of pairing is not
firmly established, which is often the case for the systems with
strong electron correlations. The standard group-theoretical
classification of the superconducting states in crystals
\cite{standard} (for a review, see Ref. \cite{min99}) is not
applicable if the normal state is ferromagnetic. In Refs.
\cite{seminal02}, the symmetry analysis was extended to cover the
magnetic case, and also a complete classification of the
superconducting states in a cubic ferromagnet such as ZrZn$_2$ was
done. The orthorhombic symmetry, which is appropriate for UGe$_2$,
was studied in Refs. \cite{uge2-theory}.

Very recently, a strong evidence was found of superconductivity
apparently co-existing with ferromagnetism in yet another class of
materials, namely graphite-sulfur composites \cite{mhm02} (see
also Ref. \cite{dasilva01}). The superconducting and ferromagnetic
behavior in highly-oriented pyrolitic graphite was reported in
Ref. \cite{kopel00}. The experimental data suggest that the
superconductivity is associated with the graphite planes, whose
symmetry is hexagonal. In this article, we study the possible
pairing symmetries and the related nodal structures of the
superconducting gap in a hexagonal ferromagnet, thus complementing
the analysis in Refs. \cite{seminal02,uge2-theory}. The details of
the electronic spectrum of graphite \cite{dress64}, the nature of
ferromagnetic and superconducting correlations, and also the fact
that its single layer has semimetallic properties with some
peculiar physics related to that \cite{graphene}, do not influence
our results. We only assume that a standard BCS-Gor'kov
phenomenology is applicable, i.e. the origin of superconductivity
is the Cooper pairing of fermionic quasiparticles with opposite
momenta, so that the order parameter is an anomalous average of
the pair creation operator.

The starting point of the group-theoretical analysis is the
symmetry group ${\cal G}$ of the normal state which is defined as
a group of transformations which leave the system Hamiltonian
invariant. In non-magnetic superconductors, time reversal symmetry
is not broken, and ${\cal G}=S\times K\times U(1)$, where $S$ is
the space group of the crystal, $K$ is time reversal operation,
and $U(1)$ is the gauge group. In contrast, in magnetic crystals
time reversal symmetry is broken, and ${\cal G}=S_M\times U(1)$,
where $S_M$ is the magnetic space group which is a group of
symmetry operations leaving both the crystal lattice (the
microscopic charge density) and the magnetization density $\bm{M}$
invariant \cite{lan60}. For example, if there is a crystal point
group rotation $R$ which transforms $\bm{M}$ to $-\bm{M}$, then
the combined operation $KR$ will be an element of $S_M$, because
the time reversal restores the original $\bm{M}$ not affecting the
lattice symmetry. We assume that the spin-orbit coupling is
sufficiently strong, so that the space group elements act on the
orbital and spin coordinates simultaneously. As shown in Refs.
\cite{magn_symm}, the symmetry analysis should be modified in the
magnetic case, due to the fact that the operation $KR$ is
anti-unitary.

The crystal structure of graphite is described by layers of
honeycomb lattices of carbon atoms, the precise arrangement of the
layers along the $z$ axis being not important for our analysis. We
make the usual assumption that the subgroup of translations (which
are replaced by magnetic translations in our case) does not affect
the momentum dependence of the order parameter. Therefore, the
space group in the paramagnetic phase can be replaced by the
hexagonal point group
$\mathbf{D}_{6h}=\mathbf{D}_6\times\mathbf{C}_i$, where
$\mathbf{C}_i=\{E,I\}$ (we assume that the space group contains
the inversion operation $I$). The group $\mathbf{D}_6$ is
generated by the rotations $C_{6z}$ about the $z$ axis by an angle
$\pi/3$, and the rotations $C_{2x}$ about the $x$ axis by an angle
$\pi$. In the ferromagnetic phase, there are three possibilities
for the easy direction of magnetization: $\bm{M}$ can be along the
$z$ axis, the $x$ axis, or the $a$ axis, where $\hat{\bm
a}=(\sqrt{3}/2,1/2,0)$. Here we study only the first possibility;
the other cases can be treated in a similar fashion. If $\bm{M}$
is along the $z$ axis, then the symmetry of the crystal is
described by the magnetic group
$\mathbf{D}_6(\mathbf{C}_6)\times\mathbf{C}_i$, which is generated
by: (i) the rotations $C_{6z}$, (ii) the operations $KC_{2x}$
which combine the rotations $C_{2x}$ with the time reversal, and
(iii) the inversion $I$. The subgroup in parentheses (the unitary
subgroup) incorporates all symmetry elements which are not
multiplied by the anti-unitary operation $KC_{2x}$, i.e.
$\mathbf{D}_6(\mathbf{C}_6)=\mathbf{C}_6+KC_{2x}\times\mathbf{C}_6$.
[If $\bm{M}$ is along the $x$ or $a$ axes, then the symmetry is
described by the magnetic group
$\mathbf{D}_2(\mathbf{C}_2)\times\mathbf{C}_i$, where
$\mathbf{D}_2(\mathbf{C}_2)=\mathbf{C}_2+KC_{2z}
\times\mathbf{C}_2$.]

In the presence of spin-orbit coupling, spin is not a good quantum
number and is replaced by pseudospin. The free electron
Hamiltonian $H_0$ in the normal state is
$H_0=\sum_{\bm{k}}[\epsilon_+(\bm{k})c^\dagger_{\bm{k}+}c_{\bm{k}+}
+\epsilon_-(\bm{k})c^\dagger_{\bm{k}-}c_{\bm{k}-}]$, where
$\epsilon_\pm(\bm{k})$ are the energy spectra for the
pseudospin-up and pseudospin-down sheets of the Fermi surface. The
transformation properties of the single-electron states
$|\bm{k},+\rangle$ and $|\bm{k},-\rangle$ with respect to the
magnetic symmetry elements are the same as those of the spin
eigenstates $|\bm{k},\uparrow\rangle$ and
$|\bm{k},\downarrow\rangle$, which means that
$C_{6z}c^\dagger_{\bm{k},\pm}C^{-1}_{6z}= e^{\mp
i\pi/6}c^\dagger_{C_{6z}\bm{k},\pm}$, $(KC_{2x})(\lambda
c^\dagger_{\bm{k},\pm})(KC_{2x})^{-1}=\pm i\lambda^*
c^\dagger_{-C_{2x}\bm{k},\pm}$, and
$Ic^\dagger_{\bm{k},\pm}I^{-1}=c^\dagger_{-\bm{k},\pm}$. Here
$\lambda$ is an arbitrary $c$-number (note that the combined
operation $KC_{2x}$ is anti-linear). The transformation properties
of the superconducting order parameter can be derived using the
mean-field approximation for the pairing Hamiltonian:
\begin{equation}
\label{H_MF}
     H_{MF}=\frac{1}{2}\sum\limits_{\bm{k}}
     \sum\limits_{\alpha,\beta=\pm}
     \Bigl[\Delta_{\alpha\beta}(\bm{k})c^\dagger_{\bm{k}\alpha}
     c^\dagger_{-\bm{k},\beta}+\mathrm{h.c.}\Bigr].
\end{equation}
From the anticommutation relations for $c_{\bm{k}\alpha}$, we see
that $\Delta_{++}(\bm{k})$ and $\Delta_{--}(\bm{k})$ are odd
functions of $\bm{k}$, but
$\Delta_{+-}(\bm{k})=-\Delta_{-+}(-\bm{k})$ does not have a
definite parity.

The order parameter matrix can be cast in a more familiar form
$\Delta(\bm{k})=(i\bm{\sigma}\sigma_2)\bm{d}(\bm{k})+
(i\sigma_2)d_0(\bm{k})$, where $\bm{d}(\bm{k})=-\bm{d}(-\bm{k})$
and $d_0(\bm{k})=d_0(-\bm{k})$ are the pseudospin-triplet and the
pseudospin-singlet components respectively \cite{min99}. It is
convenient to introduce an orthogonal basis of unit vectors
$\hat{\bm{e}}_1,\hat{\bm{e}}_2,\hat{\bm{e}}_3$ in the pseudospin
space, such that $\hat{\bm{e}}_3\parallel\bm{M}$ and use the
following representation of the vector order parameter:
$\bm{d}(\bm{k})=\hat{\bm{e}}_+d_-(\bm{k})+\hat{\bm{e}}_-d_+(\bm{k})+
\hat{\bm{e}}_3d_3(\bm{k})$, where
$\hat{\bm{e}}_\pm=(\hat{\bm{e}}_1\pm i\hat{\bm{e}}_2)/\sqrt{2}$
and $d_+=(d_1+id_2)/\sqrt{2}=\Delta_{--}/\sqrt{2}$,
$d_-=(d_1-id_2)/\sqrt{2}=-\Delta_{++}/\sqrt{2}$,
$d_3=(\Delta_{+-}+\Delta_{-+})/2$. Also,
$d_0=(\Delta_{+-}-\Delta_{-+})/2$. From Eq. (\ref{H_MF}), we
obtain the transformation rules for the order parameters under the
magnetic group operations:
\begin{eqnarray}
\label{rules_for_d}
     &&C_{6z}d_\pm(\bm{k})=e^{\pm i\pi/3}d_\pm(C^{-1}_{6z}\bm{k})
     \nonumber\\
     &&C_{6z}d_3(\bm{k})=d_3(C^{-1}_{6z}\bm{k})\nonumber\\
     &&KC_{2x}d_\pm(\bm{k})=d_\pm^*(C^{-1}_{2x}\bm{k})\\
     &&KC_{2x}d_3(\bm{k})=-d_3^*(C^{-1}_{2x}\bm{k}),\nonumber
\end{eqnarray}
and
\begin{eqnarray}
\label{rules_for_psi}
     &&C_{6z}d_0(\bm{k})=d_0(C^{-1}_{6z}\bm{k})\nonumber\\
     &&KC_{2x}d_0(\bm{k})=d_0^*(C^{-1}_{2x}\bm{k}).
\end{eqnarray}
Because of the inversion symmetry, the triplet and the singlet
order parameters can be considered separately.

\paragraph*{Triplet order parameter.}

The superconducting order parameter transforms according to one of
the irreducible representations $\Gamma$ of the normal state
symmetry group ${\cal G}$. In our case, ${\cal G}$ contains the
anti-unitary operation $KC_{2x}$, and the standard symmetry
analysis should be modified: instead of usual representations, one
should use co-representations of the magnetic point group
$\mathbf{D}_6(\mathbf{C}_6)$, which can be derived from
one-dimensional representations of the unitary subgroup
$\mathbf{C}_6$ \cite{magn_symm}. The results for odd
co-representations $\Gamma$ are listed in Table \ref{table1}. Note
that the action of the unitary and anti-unitary orbital symmetry
elements on scalar functions $f(\bm{k})$ is defined as follows:
$Rf(\bm{k})=f(R^{-1}\bm{k})$, and
$KRf(\bm{k})=f^*(-R^{-1}\bm{k})$.

The basis functions of the co-representations in Table
\ref{table1} vanish on some lines or planes in the momentum space.
Although the specific form of the basis function is not imposed by
symmetry [for example, any $f_\Gamma(\bm{k})$ can be multiplied by
an arbitrary real function of $k_z^2$, $k_x^2+k_y^2$, $(k_y\pm
ik_x)^6$, {\em etc.}], the position of the zeros in $\bm{k}$-space
is independent of the choice of $f_\Gamma(\bm{k})$. It can be
checked that the zeros of the basis functions from Table
\ref{table1} are all required by symmetry, and, on the other hand,
there are no other symmetry-imposed zeros of the basis functions.

All co-representations of $\mathbf{D}_6(\mathbf{C}_6)$ are
one-dimensional, therefore the order parameter can be represented
as
\begin{equation}
\label{d_expansion}
    \bm{d}_\Gamma(\bm{k}) = i\eta_-\hat{\bm{e}}_+f_{\Gamma_-}(\bm{k})
    +i\eta_+\hat{\bm{e}}_-f_{\Gamma_+}(\bm{k})+\eta_3
    \hat{\bm{e}}_3f_{\Gamma_3}(\bm{k}).
\end{equation}
Thus, the order parameter has three components $\eta_+$, $\eta_-$,
and $\eta_3$, which enter the Ginzburg-Landau (GL) free energy and
can be $\bm{r}$-dependent, in general. It is obvious from Eqs.
(\ref{rules_for_d}) and (\ref{d_expansion}) that the orbital
symmetries of $d_+$, $d_-$, and $d_3$, which are described by the
co-representations $\Gamma_+$, $\Gamma_-$, and $\Gamma_3$
respectively, should all be different. In Table \ref{table2}, we
list the combinations of the orbital co-representations which give
rise to the same symmetry $\Gamma$ of the vector $\bm{d}$. For
example, the vector
$\bm{d}(\bm{k})=i\eta_-\hat{\bm{e}}_+f_{B_u}(\bm{k})+
i\eta_+\hat{\bm{e}}_-f_{{}^1E_{1u}}(\bm{k})+
\eta_3\hat{\bm{e}}_3f_{{}^2E_{2u}}(\bm{k})$ transforms according
to the co-representation ${}^2E_{2u}$.

The transformations of the pseudospin vector $\bm{d}$ under the
magnetic symmetry elements can be interpreted as operations acting
on $\bm{\eta}=(\eta_+,\eta_-,\eta_3)^T$. Namely, for a
co-representation $\Gamma$,
$C_{6z}\bm{\eta}(\bm{r})=\chi_\Gamma[C_{6z}]\bm{\eta}(\bm{r})$,
where $\chi_\Gamma$ is the character of $C_{6z}$ from Table
\ref{table1}. Also, because of our choice of the overall phases of
the basis functions (see the caption to Table \ref{table1}) and
the presence of the factors $i$ on the right-hand side of Eqs.
(\ref{d_expansion}),
$KC_{2x}\bm{\eta}(\bm{r})=\bm{\eta}^*(\bm{r})$. Note that there
are no symmetry operations that transform, say, $\eta_+$ to
$\eta_-$, {\em etc}, which has profound consequences for the GL
theory of ferromagnetic superconductors. Indeed, the GL free
energy contains all combinations of the order parameter components
and the spatial gradients that are invariant under the
transformations from ${\cal G}$. The fact that the only change
that occurs to $\bm{\eta}$ under the magnetic symmetry operations
is the multiplication by a phase factor which is the same for all
three components means that there are more invariant terms allowed
in the GL functional for a ferromagnetic superconductor than for
its non-ferromagnetic counterpart. The general form of the uniform
terms in the free energy density is
\begin{equation}
\label{GL_uniform}
   F = \sum\limits_{ij=\pm,3}A_{ij}(T)\eta_i^*\eta_j+
   \sum\limits_{ijkl=\pm,3}B_{ij,kl}\eta_i^*\eta_j^*\eta_k\eta_l,
\end{equation}
where $A_{ij}$ is a real symmetric matrix, and the matrix $B$ is
symmetric with respect to $i\leftrightarrow j$ and
$k\leftrightarrow l$, and satisfies the following condition:
$B_{ij,kl}=B_{kl,ij}$. The critical temperature $T_c$ is defined
as the temperature at which one of the eigenvalues of $A$ changes
sign. At $T>T_c$, $A$ is positive definite, and
$\eta_+=\eta_-=\eta_3=0$. Below $T_c$, all three components of
$\bm{\eta}$ are non-zero, in general. In addition to Eq.
(\ref{GL_uniform}), the GL functional contains a variety of the
gradient terms. All this can lead to a rich phase diagram, which
we shall not discuss here. Let us just note Eq. (\ref{GL_uniform})
is formally equivalent to a model of a three-band superconductor
with three scalar order parameters of the same symmetry.

\begin{table}
\caption{\label{table1} The character table and the examples of
the basis functions for the odd irreducible co-representations of
the magnetic point group $\mathbf{D}_6(\mathbf{C}_6)$. The overall
phases of the basis functions are chosen so that
$KC_{2x}f_\Gamma(\bm{k})=f_\Gamma(\bm{k})$. $\omega=\exp(2\pi
i/3)$, and $\lambda_{1,2}$ are arbitrary real constants.}
\begin{ruledtabular}
\begin{tabular}{|c||c|c|c|}
   $\Gamma$  & $E$ & $C_{6z}$ &  $f_\Gamma(\bm{k})$ \\ \hline
   $A_u$      & 1    & 1  &  $k_z$  \\ \hline
   $B_u$      & 1    & $-1$ &  $\lambda_1(k_y+ik_x)^3+
            \lambda_2(k_y-ik_x)^3$\\\hline
   ${}^1E_{1u}$  & 1    & $-\omega^*$ &  $k_y+ik_x$ \\ \hline
   ${}^2E_{1u}$  & 1    & $-\omega$ &  $k_y-ik_x$ \\ \hline
   ${}^1E_{2u}$  & 1    & $\omega^*$ &  $k_z(k_y-ik_x)^2$ \\ \hline
   ${}^2E_{2u}$  & 1    & $\omega$ &  $k_z(k_y+ik_x)^2$
\end{tabular}
\end{ruledtabular}
\end{table}

\begin{table}
\caption{\label{table2} The sets of orbital co-representations
corresponding to the same symmetry of the triplet order parameter
$\bm{d}_\Gamma(\bm{k})$ [see Eq. (\ref{d_expansion})].}
\begin{ruledtabular}
\begin{tabular}{|c||c|}
   $\Gamma$  & $(f_{\Gamma_+}(\bm{k}),f_{\Gamma_-}(\bm{k}),
                 f_{\Gamma_3}(\bm{k}))$ \\ \hline
   $A_u$      & $({}^2E_{1u}, {}^1E_{1u}, A_u)$  \\ \hline
   $B_u$      & $({}^2E_{2u}, {}^1E_{2u}, B_u)$  \\ \hline
   $^1E_{1u}$  & $(A_u, {}^2E_{2u}, {}^1E_{1u})$ \\ \hline
   $^2E_{1u}$  & $({}^1E_{2u}, A_u, {}^2E_{1u})$ \\ \hline
   $^1E_{2u}$  & $(B_u, {}^2E_{1u}, {}^1E_{2u})$ \\ \hline
   $^2E_{2u}$  & $({}^1E_{1u}, B_u, {}^2E_{2u})$
\end{tabular}
\end{ruledtabular}
\end{table}

An important consequence of the above results is that the gap in
the spectrum of Bogoliubov quasiparticles goes to zero at some
points or lines at the Fermi surface. The excitation spectrum can
be obtained by diagonalizing the Hamiltonian $H=H_0+H_{MF}$. The
quasiparticle energy $E(\bm{k})$ vanishes at some $\bm{k}$ if the
following condition is satisfied:
\begin{eqnarray}
\label{EB}
    \epsilon_+^2\epsilon_-^2+2\epsilon_+^2|d_+|^2 +
    2\epsilon_-^2|d_-|^2+2\epsilon_+\epsilon_-|d_3|^2\nonumber\\
    +|2d_+d_-+d_3^2|^2 = 0.
\end{eqnarray}
Thus, the condition for the gap zeros on the pseudospin-up sheet
of the Fermi surface [at $\epsilon_+(\bm{k})=0$], is that
$d_-(\bm{k})=d_3(\bm{k})=0$, i.e.
$f_{\Gamma_-}(\bm{k})=f_{\Gamma_3}(\bm{k})=0$. For the gap zeros
on the pseudospin-down sheet [at $\epsilon_-(\bm{k})=0$], we must
have $d_+(\bm{k})=d_3(\bm{k})=0$, i.e.
$f_{\Gamma_+}(\bm{k})=f_{\Gamma_3}(\bm{k})=0$. Using Tables
\ref{table1} and \ref{table2}, we see that if the symmetry of
$\bm{d}$ corresponds to the co-representations $B_u$,
${}^1E_{2u}$, or ${}^2E_{2u}$, then all three orbital basis
functions have zeros on the line $k_x=k_y=0$, so that the energy
gap vanishes at the poles of both sheets of the Fermi surface. On
the contrary, for the ${}^1E_{1u}$ and ${}^2E_{1u}$ symmetries,
the gap goes to zero at the poles of one of the sheets, while the
other sheet remains fully gapped. For the $A_u$ symmetry, there
are no symmetry-imposed gap zeros.

In the discussion above, we implicitly assumed that all three
components of $\bm{d}$ have comparable magnitude. More realistic
scenario is that the conditions for the appearance of
superconductivity are more favorable on one of the sheets of the
Fermi surface, so that only one component of the order parameter,
say $d_-$ on the pseudospin-up sheet, is dominant. Then, the $d_+$
component is induced by the inter-band interactions of the form
$c^\dagger_{\bm{k}+}c^\dagger_{-\bm{k},+}c_{\bm{k}'-}c_{-\bm{k}',-}$,
which are expected to be small if the spin-orbit coupling is weak
(these interactions vanish at zero spin-orbit coupling because of
the spin conservation). Also, if the exchange band splitting
greatly exceeds the superconducting energy scale, which is of the
order of $T_c$, then the inter-band interactions
$c^\dagger_{\bm{k}+}c^\dagger_{-\bm{k},-}c_{\bm{k}'-}c_{-\bm{k}',+}$
responsible for the $d_3$-component, are negligibly small. While
this is the case in such materials as ZrZn$_2$ and UGe$_2$, it is
not clear whether it is also true for the graphite-based
ferromagnets. If the $d_3$-component can indeed be neglected, then
the conditions for the presence of the gap zeros become less
restrictive, so that, as seen from Tables \ref{table1} and
\ref{table2}, the energy gap always has either line and/or point
nodes on both sheets of the Fermi surface.

\paragraph*{Singlet order parameter.}

Similar to the $d_3$-component of the triplet pairing, the singlet
pairing can only be realized if the exchange band splitting is
less than the superconducting $T_c$ (Chandrasekhar-Clogston limit)
\cite{cc_limit64}. This limit can be slightly exceeded if to
consider the possibility of the Cooper pairs having a non-zero
momentum (Larkin-Ovchinnikov-Fulde-Ferrell state) \cite{loff}. The
symmetry analysis can be done similarly to the triplet case. The
only difference is that there is only one order parameter
component, and the symmetry is described by even
co-representations of $\mathbf{D}_6(\mathbf{C}_6)$. We have
\begin{equation} \label{psi_expansion}
    d_{0,\Gamma}(\bm{k}) = \psi f_\Gamma(\bm{k}),
\end{equation}
where $\psi$ is a quantity which enters the GL functional, and
$f_\Gamma(\bm{k})$ is the basis function for the co-representation
$\Gamma$, see Table \ref{table3}. As mentioned above,
$f_\Gamma(\bm{k})$ can be multiplied by an arbitrary real function
of $k_z^2$, $k_x^2+k_y^2$, $(k_y\pm ik_x)^6$, {\em etc}. Under the
action of the magnetic symmetry elements,
$C_{6z}\psi(\bm{r})=\chi_\Gamma[C_{6z}]\psi(\bm{r})$, where
$\chi_\Gamma$ is the character of $C_{6z}$ from Table
\ref{table3}, and also $KC_{2x}\psi(\bm{r})=\psi^*(\bm{r})$.

\begin{table}
\caption{\label{table3} The character table and the examples of
the basis functions for the even irreducible co-representations of
the magnetic point group $\mathbf{D}_6(\mathbf{C}_6)$. For all
$\Gamma$, $KC_{2x}f_\Gamma(\bm{k})=f_\Gamma(\bm{k})$, and
$\lambda_{1,2}$ are arbitrary real constants.}
\begin{ruledtabular}
\begin{tabular}{|c||c|c|c|}
   $\Gamma$  & $E$ & $C_{6z}$ &  $f_\Gamma(\bm{k})$ \\ \hline
   $A_g$      & 1    & 1  &  $1$  \\ \hline
   $B_g$      & 1    & $-1$ &  $k_z[\lambda_1(k_y+ik_x)^3+
            \lambda_2(k_y-ik_x)^3]$\\\hline
   ${}^1E_{1g}$  & 1    & $-\omega^*$ &  $k_z(k_y+ik_x)$ \\ \hline
   ${}^2E_{1g}$  & 1    & $-\omega$ &  $k_z(k_y-ik_x)$ \\ \hline
   ${}^1E_{2g}$  & 1    & $\omega^*$ &  $(k_y-ik_x)^2$ \\ \hline
   ${}^2E_{2g}$  & 1    & $\omega$ &  $(k_y+ik_x)^2$
\end{tabular}
\end{ruledtabular}
\end{table}

The condition for the gap in the excitation energy $E(\bm{k})$ to
vanish at some $\bm{k}$ is simply
\begin{equation}
    \epsilon_+\epsilon_-+|d_0|^2=0,
\end{equation}
therefore the gap nodes appear simultaneously on both sheets of
the Fermi surface where the basis function $f_\Gamma(\bm{k})$ has
symmetry-imposed zeros. From Table \ref{table3}, we see that for
all order parameter symmetries, except from $A_g$, the gap goes to
zero either on the equators or at the poles of both Fermi
surfaces.

The expressions for the basis functions of the co-representations
of $\mathbf{D}_6(\mathbf{C}_6)$ given in Tables \ref{table1} and
\ref{table3} are applicable if the Fermi surface is centered
around the $\Gamma$ point of the first Brillouin zone. It is not
the case in graphite, where the Fermi surface consists of small
``sausage''-like pockets along the six vertical edges of the
hexagonal Brillouin zone, i.e. along the lines
$\bm{k}=C^n_{6z}\bm{k}_s$, where $n=0,...,5$,
$\bm{k}_s=2\bm{K}_1/3-\bm{K}_2/3+k_z\hat{\bm{e}}_z$, and
$\bm{K}_1$ and $\bm{K}_2$ are the reciprocal lattice vectors
\cite{dress64}. It can be easily proved that for all odd and even
$E$ co-representations, in addition to the zeros at $k_x=k_y=0$,
the basis functions also vanish at the vertical edges of the
Brillouin zone, because $\bm{k}_s$ and $C_{3z}\bm{k}_s$ are
equivalent points. Similarly, if a basis function vanishes on the
plane $k_z=0$, then it should also vanish at the horizontal
surfaces of the Brillouin zone, i.e. at $k_z=\pm\pi/c_0$, where
$c_0$ is the lattice constant of graphite along the $z$ axis.

The presence of the gap nodes would manifest itself in power-law
temperature dependences of the thermodynamic and transport
properties \cite{min99}. For example, the electronic specific heat
at low temperatures should be $C(T)/T\sim T^2$ for the first-order
point nodes, $C(T)/T\sim T$ for the line nodes or the second-order
point nodes, and $C(T)/T\sim T^{2/3}$ for the third-order point
nodes.

To summarize, we have derived a complete symmetry classification
of the superconducting states in a hexagonal ferromagnet,
considering both the triplet and the singlet pairing channels. We
have discussed the nodal structure of the gap in the excitation
spectrum, and also the modifications to the Ginzburg-Landau theory
for ferromagnetic superconductors. So far, the only experimental
system to which our results might be applicable is the
graphite-sulfur composite studied in Ref. \cite{mhm02}.

The author has greatly benefited from the numerous discussions
with M.~Walker about the symmetry of ferromagnetic
superconductors. The author thanks B. Mitrovic for useful comments
and interest to this work, and also F.~Razavi and M.~Reedyk for
the discussions of the experimental situation. The financial
support from Brock University is gratefully acknowledged.


\begin{thebibliography}{99}

\bibitem{zrzn2-exp}
C. Pfleiderer, M. Uhlarz, S. M. Hayden, R. Vollmer,
H.~v.~L\"{o}hneysen, N. R. Bernhoeft, and G. G. Lonzarich,
Nature {\bf 412}, 58 (2001).

\bibitem{uge2-exp}
S. S. Saxena, P. Agarwal, K. Ahilan, F. M. Grosche,
R.~K.~W.~Haselwimmer, M. J. Steiner, E. Pugh, I.~R.~Walker, S. R.
Julian, P. Monthoux, G. G. Lonzarich, A. Huxley, I. Sheikin, D.
Braithwaite, and J. Flouquet, Nature {\bf 406}, 587 (2000).

\bibitem{standard}
G. E. Volovik and L. P. Gor'kov, Sov. Phys. -- JETP {\bf 61}, 843
(1985) [Zh. Eksp. Teor. Fiz. {\bf 88}, 1412 (1985)]; K.~Ueda and
T. M. Rice, Phys. Rev. B {\bf 31}, 7114 (1985); E.~I.~Blount,
Phys. Rev. B {\bf 32}, 2935 (1985).

\bibitem{min99}
V. P. Mineev and K. V. Samokhin, {\it Introduction to
Unconventional Superconductivity} (Gordon and Breach, The
Netherlands, 1999).

\bibitem{seminal02}
M. B. Walker and K. V. Samokhin, Phys. Rev. Lett. {\bf 88}, 207001
(2002); K. V. Samokhin and M. B. Walker, Phys. Rev. B {\bf 66},
024512 (2002); K. V. Samokhin and M.~B.~Walker,
arXiv:cond-mat/0206487.

\bibitem{uge2-theory}
V. P. Mineev, arXiv:cond-mat/0204263; I. A. Fomin,
arXiv:cond-mat/0207152 (see also I. A. Fomin, JETP Lett. {\bf 74},
111 (2001) [Pis'ma Zh. Eksp. Teor. Fiz. {\bf 74}, 116 (2001)], for
an earlier attempt of the symmetry analysis for UGe$_2$).

\bibitem{mhm02}
S. Moehlecke, P.-C. Ho, and M. B. Maple, arXiv:cond-mat/0204006.

\bibitem{dasilva01}
R. Ricardo da Silva, J. H. S. Torres, and Y. Kopelevich, Phys.
Rev. Lett. {\bf 87}, 147001 (2001).

\bibitem{kopel00}
Y. Kopelevich, P. Esquinazi, J. H. S. Torres, and S.~Moehlecke, J.
Low Temp. Phys. {\bf 119}, 691 (2000).

\bibitem{dress64}
M. S. Dresselhaus and J. G. Mavroides, IBM J. Res. Develop. {\bf
8}, 262 (1964); S. J. Williamson, S. Foner, and M.~S.~Dresselhaus,
Phys. Rev. {\bf 140}, A1429 (1965).

\bibitem{graphene}
J. Gonzalez, F. Guinea, and M. A. H. Vozmediano, Phys. Rev. B {\bf
63}, 134421 (2001); D. V. Khveshchenko, Phys. Rev. Lett. {\bf 87},
246802 (2001); G. Baskaran and S.~A.~Jafari, Phys. Rev. Lett. {\bf
89}, 016402 (2002).

\bibitem{lan60}
L. D. Landau and E. M. Lifshitz, {\it Electrodynamics of
Continuous Media} (Pergamon Press, London, 1960).

\bibitem{magn_symm}
E.~Wigner, {\it Group Theory and Its Applications to the Quantum
Mechanics of Atomic Spectra} (Academic Press, NY, 1959);
C.~J.~Bradley and A.~P.~Cracknell,
{\it The Mathematical Theory of Symmetry in Solids}
(Clarendon Press, Oxford, 1972);
A.~P.~Cracknell, Progr. Theor. Phys. {\bf 35}, 196 (1966).

\bibitem{cc_limit64}
B. S. Chandrasekhar, Appl. Phys. Lett. {\bf 1}, 7 (1962);
A.~M.~Clogston, Phys. Rev. Lett. {\bf 9}, 266 (1962).

\bibitem{loff}
A. I. Larkin and Yu. N. Ovchinnikov, Sov. Phys. -- JETP {\bf 20},
762 (1965) [Zh. Eksp. Teor. Fiz. {\bf 47}, 1136 (1964)]; P.~Fulde
and R. A. Ferrell, Phys. Rev. {\bf 135}, 550 (1964).



\end{thebibliography}
\end{document}